%% file: Deformed-GNR.tex
\documentclass[prb,aps,amstex,preprint,amsmath,amssymb,array,tabularx,nofootinbib]{revtex4}
\usepackage{graphicx}
\usepackage{dcolumn}
\usepackage{bm}
\usepackage{lscape} 
\usepackage{amsmath}
\usepackage{amsxtra}
\usepackage{tabularx}
\usepackage{footmisc}

\DeclareGraphicsExtensions{.jpg,.pdf,.eps,.gif}

\begin{document}

\title{Electromechanical properties of suspended Graphene Nanoribbons}

\author{$\mbox{Oded Hod}$}

\affiliation{School of Chemistry, The Sackler Faculty of Exact Sciences,
Tel Aviv University, Tel Aviv 69978, Israel}

\author{$\mbox{Gustavo E. Scuseria}$}

\affiliation{Department of Chemistry, Rice University, Houston,
  Texas 77005-1892}

\date{\today}

\begin{abstract}
  \input{Abstract}
\end{abstract}

\maketitle
 


\renewcommand{\thefootnote}{\fnsymbol{footnote}}

\input{Intro}

\input{Mech}

\input{ElectMech}

\input{Summary}

\bibliographystyle{prsty} \bibliography{Deformed-GNR}
\end{document}

%% file: Abstract.tex
Graphene nanoribbons present diverse electronic properties ranging
from semi-conducting~\cite{Ezawa2006, Barone2006, Son2006-2} to
half-metallic,~\cite{Son2006} depending on their geometry, dimensions
and chemical composition.~\cite{Hod2007-I} Here we present a route to
control these properties via externally applied mechanical
deformations. Using state-of-the-art density functional theory
calculations combined with classical elasticity theory considerations,
we find a remarkable Young's modulus value of $\sim 7$~TPa for
ultra-narrow graphene strips and a pronounced electromechanical
response towards bending and torsional deformations. Given the current
advances in the synthesis of nanoscale graphene derivatives, our
predictions can be experimentally verified opening the way to the
design and fabrication of miniature electromechanical sensors and
devices based on ultra-narrow graphene nanoribbons.


%% file: Intro.tex
\begin{figure}[h]
\begin{center}
\epsfxsize=14.0cm \epsffile{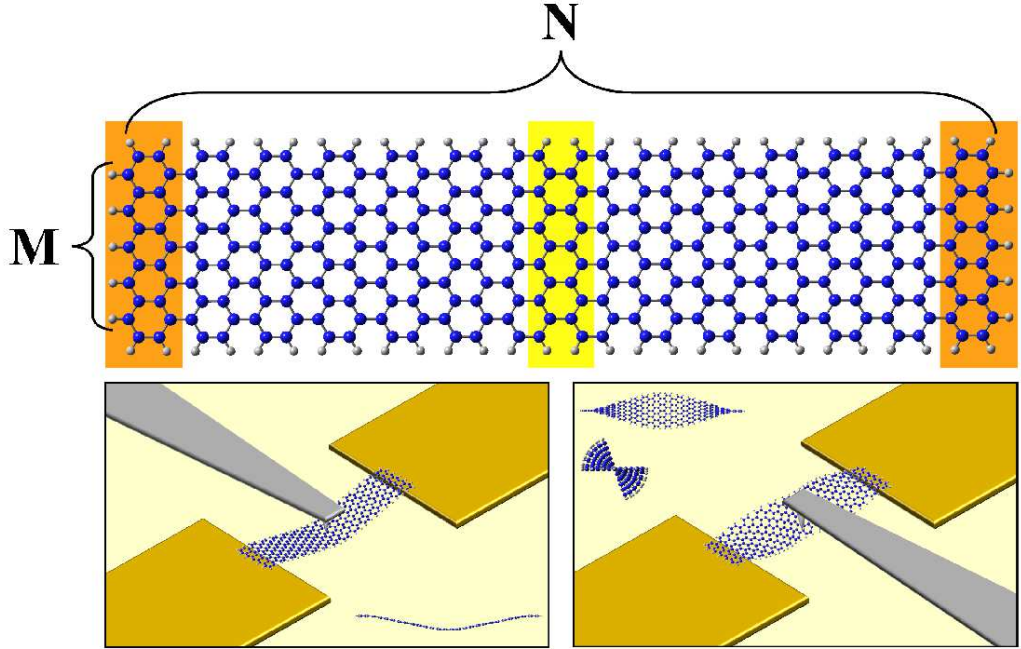}
\end{center}
\caption{Illustration of a graphene nanoribbon based
  electromechanical device.  Upper panel: a representative structure
  of the graphene nanoribbons set studied.  The notation $M$x$N$ is
  used to represent a GNR with M hydrogen atoms passivating each
  zigzag edge and N hydrogen atoms passivating the armchair edge. The
  deformations are simulated by clamping the atomic strips at the
  zigzag edges (highlighted by orange rectangles) and depressing or
  rotating the atoms at the center of the ribbon (highlighted by a
  yellow rectangle) with respect to the fixed edges. Lower panels: An
  artist view of a suspended GNR under bending (lower left panel) and
  torsional (lower right panel) deformation induced by an external
  tip. Side views of the deformed structures are presented for clarity
  as insets.}
\label{Fig: Ribbon}
\end{figure}

The promise of graphene nanoribbons (GNRs) as ultimate building blocks
for future nanoelectromechanical systems (NEMS) has been recently
demonstrated experimentally for the first time.~\cite{Bunch2007} Since
their initial successful fabrication,~\cite{Geim2004} the dimensions
of GNRs have rapidly reduced from the microscale down to a record
breaking value of a few nanometers in width, fabricated by either
top-down~\cite{Li2008, Wang2008} or bottom-up~\cite{Yang2008}
approaches.  While the electromechanical characteristics of their
cylindrical counterparts, carbon nanotubes (CNTs), have attracted
great interest in recent years,~\cite{Tombler2000, Rueckes2000,
Maiti2003, Minot2003, Sazonova2004, Navarro2004, Semet2005, Karni2006,
Stampfer2006, Hall2007} similar effects in GNRs have been hardly
addressed,~\cite{Bunch2007, Poot2008} thus far.  Here, we study the
electromechanical response of GNRs by considering a large set of
hydrogen-terminated GNRs with varying lengths and widths. The effect
of uniaxial strain on the electronic properties of infinite GNRs has
been recently studied in detail.~\cite{Sun2008} To simulate the {\em
bending} and {\em torsion} deformations of a finite suspended
GNR~\cite{Bunch2007} we clamp the atoms in the region close to the
zigzag edges (highlighted by orange rectangles in the upper panel of
Fig.~\ref{Fig: Ribbon}) while applying the deformation to the atoms
residing within a narrow strip along its center-line (highlighted by a
yellow rectangle in the figure).  For bending deformations the central
atomic strip is depressed down with respect to the fixed edges
parallel to its original location in the plane of the unperturbed
nanoribbon (lower left panel of Fig.~\ref{Fig: Ribbon}). Torsional
deformations are simulated by rotating the central atomic strip around
the main axis of the GNR and fixing its location with respect to the
plane of the clamped edge atoms (lower right panel of Fig.~\ref{Fig:
Ribbon}).  The positions of the remaining atoms are relaxed using
spin-polarized density functional theory calculations within the
screened-exchange hybrid approximation of Heyd, Scuseria, and
Ernzerhof (HSE06),~\footnote[2]{The HOMO-LUMO gap energy is of central
importance in the present study. Therefore, we use the HSE06
functional, which accurately predicts bandgap values for
GNRs.~\cite{Han2007, Chen2007} Values calculated using the local
density or the gradient corrected approximation usually underestimate
GNRs bandgaps when compared to experimental
measurements.}$^,$~\cite{Heyd2003, Heyd2006} as implemented in the
development version of the {\it Gaussian} suit of
programs,~\cite{gdv_short} and the $\mbox{double-}\zeta$ polarized
$\mbox{6-31G}^{**}$ Gaussian basis set.~\cite{Hariharan1973} This
theoretical approach has been recently shown to describe the physical
properties of graphene based materials with exceptional
success.~\cite{Barone2005-I, Barone2005-II, Barone2006, Hod2007-I,
Hod2007-II, Hod2008-I, Hod2008-II}
\input{epsf}

%% file: Mech.tex
\begin{figure}[h]
\begin{center}
\epsfxsize=14.0cm \epsffile{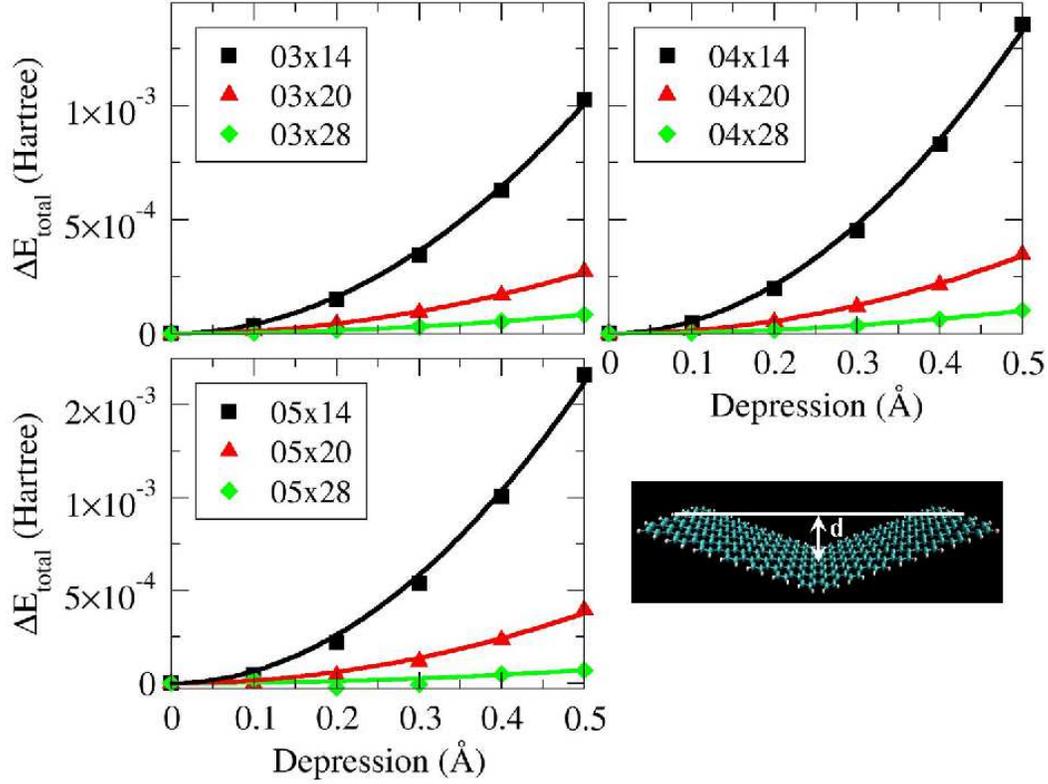}
\end{center}
\caption{ Total energy changes of the $03$x$N$ (upper left panel), $04$x$N$ (upper
right panel), and $05$x$N$ (lower left panel) suspended graphene
nanoribbons due to an externally applied bending stress. Marks
represent calculated results. Lines are parabolic fits indicating that
all systems are within the linear response regime. Lower right panel:
the depression depth is defined as the distance between the lowered
central atomic strip and the original plane of the unconstrained GNR.}
\label{Fig: EtotBend}
\end{figure}
\begin{figure}[h]
\begin{center}
\epsfxsize=14.0cm \epsffile{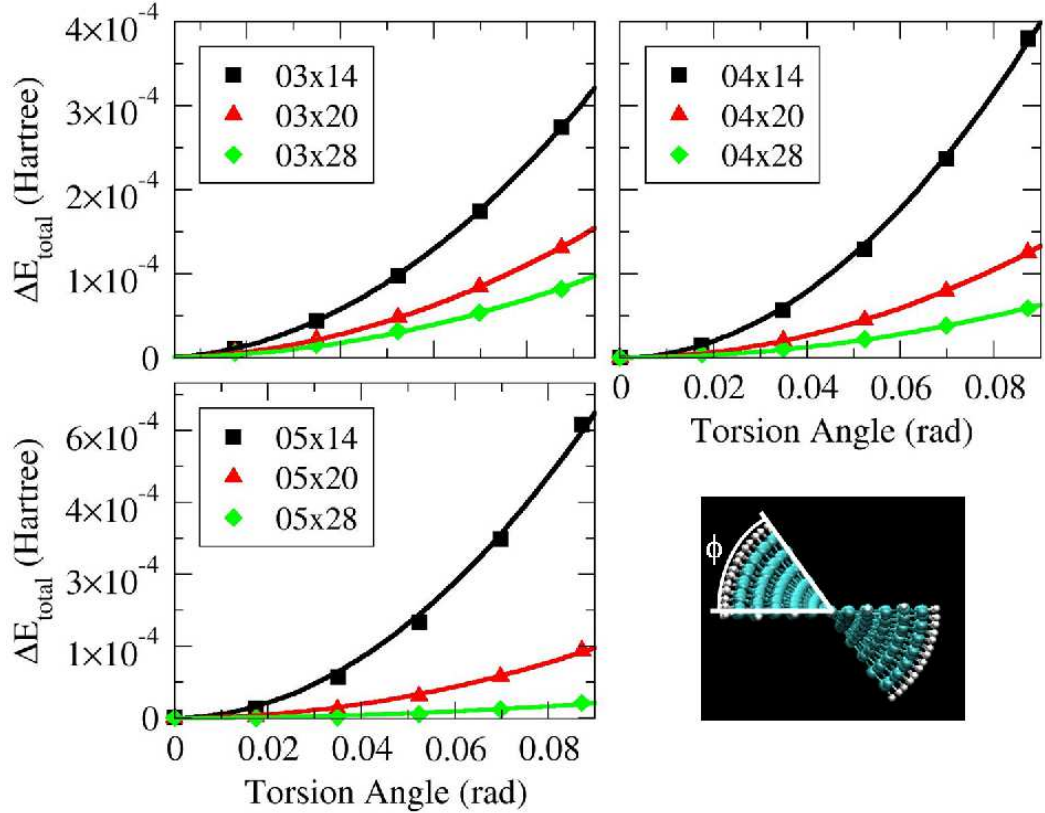}
\end{center}
\caption{Total energy changes of the $03$x$N$ (upper left panel), $04$x$N$
(upper right panel), and $05$x$N$ (lower left panel) suspended
graphene nanoribbons due to an externally applied torsional
stress. Marks represent calculated results. Lines are parabolic fits
indicating that all systems are within the linear response
regime. Lower right panel: the torsional angle is defined as the
angle between the rotated central atomic strip and the original
plane of the unconstrained GNR.}
\label{Fig: EtotTwist}
\end{figure}

To understand the mechanical response of GNRs under an externally
applied stress, macroscopic elasticity theory concepts are here
adopted.  We quantify the bending deformation by defining the
depression depth, $d$, as the distance of the central atomic strip
from the original plane of the unconstrained nanoribbon (see lower
right panel of Fig.~\ref{Fig: EtotBend}). Similarly, the torsional
angle is defined as the angle between the plane of the central atomic
strip and the plane defined by the clamped edge atoms (see lower right
panel of Fig.~\ref{Fig: EtotTwist}). In Fig.~\ref{Fig: EtotBend}, the
change in total energy as a function of the depression depth within
the linear response regime is presented for the full set of GNRs
studied.  The marks represent calculated values while the lines are
parabolic fits, from which the bending force constants are extracted.
\input{epsf}
The following simple relation describes the dependence of the bending
force constant of a macroscopic rectangular rod on its
dimensions:~\cite{Venkatesh2005}
\begin{equation}
K_b=16Yw\left(\frac{t}{l}\right)^3,
\label{Eq: Bending}
\end{equation}
where $w$ is the width of the rod, $t$ the thickness of the rod, $l$
the length of the rod, and $Y$ its Young's modulus. We now fit the
bending force constants calculated via the parabolic curves in
Fig.~\ref{Fig: EtotBend} to this simple relation. The length of the
sample is taken to be the minimum distance between fixed edge atoms on
the opposing clamped edges of the GNR. The width is taken as the
minimum distance between hydrogen atoms passivating opposite armchair
edges. For the thickness, we assume a typical value of $t=0.75$ \AA
\mbox{ }for the graphene sheet~\cite{Huang2006}.  In Fig~\ref{Fig:
Young} the dependence of the bending force constant on the inverse
cube of the length (left panel) and on the width (right panel) of the
ribbon are presented.  We find excellent correlation between our
calculated values and predictions from elasticity theory.  The Young's
modulus of GNRs can be extracted from the slopes of the linear
curves. An impressive large value of $\sim 7$ TPa is obtained,
exceeding the measured value for micrometer scale suspended graphene
sheets~\cite{Lee2008} and the highest value calculated for CNTs with
similar thickness parameters.~\cite{Huang2006}
\input{epsf}
\begin{figure}[h]
\begin{center}
\epsfxsize=14.0cm \epsffile{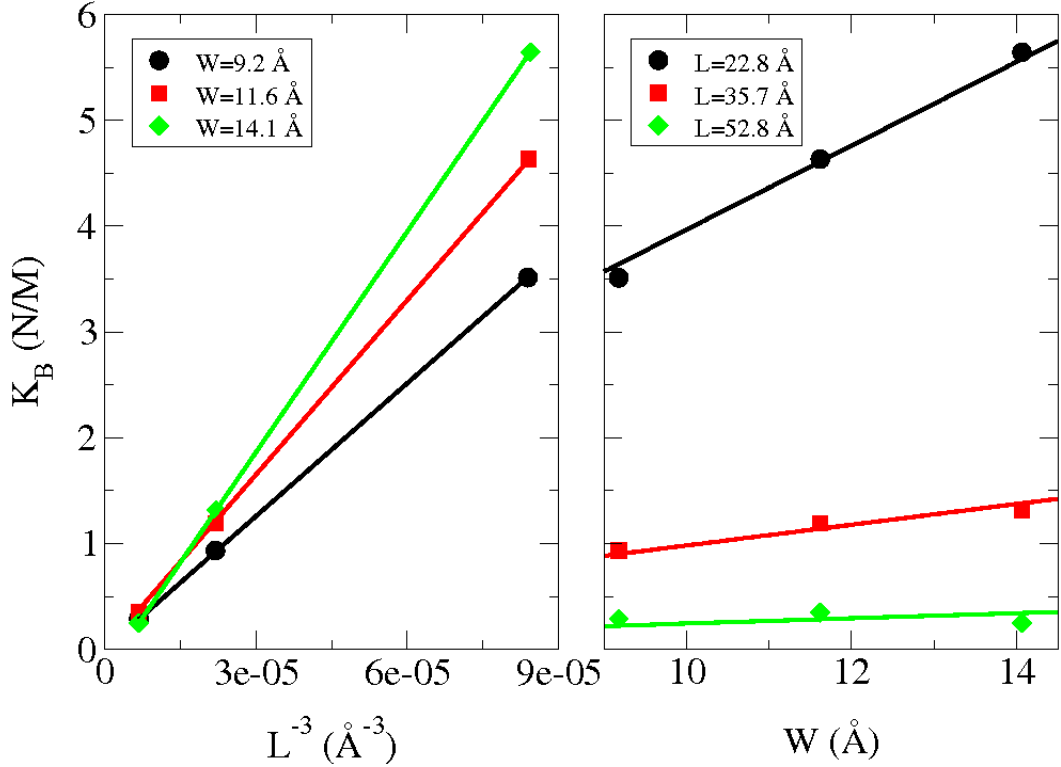}
\end{center}
\caption{The dependence of the bending force constant on the inverse cube of
the length at constant width (left panel) and of the width at constant
length (right panel) of the GNRs studied. A Young's modulus value of
$\sim 7$ TPa is calculated from the slopes of the linear curves by the
use of Eq.~\ref{Eq: Bending}.}
\label{Fig: Young}
\end{figure}

A similar analysis can be performed for torsional deformations.  In
Fig.~\ref{Fig: EtotTwist}, the change in total energy as a function of
the torsional angle within the linear response regime is presented for
the full set of GNRs studied.  As before, the marks represent
calculated values and the lines are parabolic fits.  While, similar to
the case of bending deformations, excellent parabolic fits are
obtained, classical theory of elasticity fails to predict the behavior
of the torsional force constant with the dimensions of the system and
hence it is impossible to extract the shear modulus of GNRs from these
calculations.
\input{epsf}
Several factors may limit the validity of the classical theory for the
description of torsional deformations in the considered molecular
systems. The poorly defined thickness leading to extremely large
surface to volume ratios combined with the overall nanoscopic
dimensions of the systems suggest that the equations should be
quantized in order to predict the correct behavior. Furthermore, the
effect of elongation during the twisting process may introduce
considerable deviations from the pure torsional equations.  Therefore,
it is impressive that for the case of bending deformations the effect
of these factors is small and very good agreement between the behavior
of a macroscopic rod and that of a nanoscale atomic sheet is obtained.

%% file: ElectMech.tex
Having explored the mechanical properties of suspended GNRs, we now
turn to evaluate their electromechanical responses. For this purpose,
we consider large bending and torsional deformations, well beyond the
linear response regime. Remarkably, we find that most of the studied
systems can sustain extreme mechanical deformations while retaining
their elastic nature. The HOMO-LUMO gap, which is the difference
between the highest-occupied and lowest-unoccupied molecular orbital
energies, is used to evaluate the influence of the mechanical
deformations on the electronic properties of the system.
\input{epsf}
\begin{figure}[h]
\begin{center}
\epsfxsize=14.0cm \epsffile{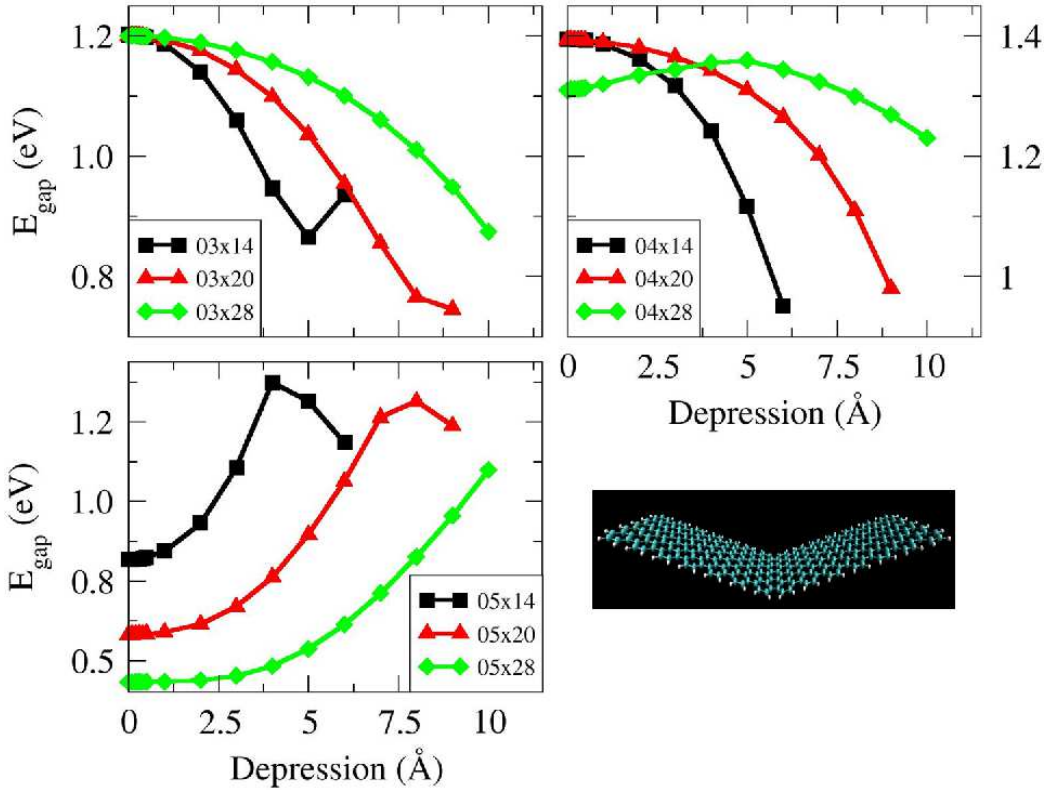}
\end{center}
\caption{HOMO-LUMO gap changes of the $03$x$N$ (upper left panel), $04$x$N$
(upper right panel), and $05$x$N$ (lower left panel) suspended
graphene nanoribbons due to an externally applied bending stress.}
\label{Fig: GapBending}
\end{figure}
In Fig.~\ref{Fig: GapBending} the energy gap as a function of the
depression depth is presented for the full set of GNRs studied. High
sensitivity of the energy gap upon the depression depth is found. For
most of the systems studied, the gap energy changes by $\sim 0.5$ eV
upon a depression of $1$ nm.  As expected, the smaller the system
dimensions, the higher the sensitivity upon similar stress
conditions. Interestingly, the $03$x$N$ and the narrower $04$x$N$
group members show an initial decrease in the energy gap upon bending
while the $05$x$N$ group presents an initial increase.  This resembles
the case of CNTs, where strain induced bandgap dependence was
attributed to Brillouin zone deformations upon the development of
mechanical stresses in the system.~\cite{Yang2000} If we regard the
finite GNRs as unrolled segments of zigzag CNTs, and follow the theory
of Yang and Han,~\cite{Yang2000} we expect to see minor bandgap
changes due to bending. Nevertheless, because of the doubly-clamped
geometry, the bending is not pure and includes a considerable
stretching component.  In the case of $(n,0)$ zigzag CNTs, the sign of
the bandgap change upon uniaxial stretching is~\cite{Yang2000}
$sgn(\Delta E_{gap})=sgn(2p+1)$, where $n=3q+p$, $q$ being an integer
number, and $p=0,\pm 1$. Therefore, the bandgap of metallic CNTs
($p=0$) is expected to increase upon uniaxial stretching.  This is the
case for the $05$x$N$ group, which can be regarded as an unrolled
$(6,0)$ CNT and in the limit of $n\rightarrow\infty$ becomes nearly
metallic.~\cite{Barone2006, Hod2008-I} When regarding the $04$x$N$
systems as unrolled $(5,0)$ CNTs, which are characterized by $p=-1$,
it is expected that the energy gap will decrease upon deformations
which involve bond stretching, as is the case for the $04$x$14$ and
$04$x$20$ nanoribbons.  Interestingly, deviations from this role occur
for the $04$x$28$ and the $03$x$N$ group members, which correspond to
an unrolled $(4,0)$ CNT with $p=1$. Such deviations are expected due
to the fact that the systems under consideration are of extremely
small dimensions, where the relevance of the band-structure theory,
upon which the Yang-Han theory is based, is limited. Furthermore, the
Born-Von K\'{a}rm\'{a}n boundary conditions in the circumferential
direction of CNTs, taken into account in the Yang-Han theory are
replaced by particle-in-a-box like boundary conditions in GNRs, which
may change the overall behavior of the bandgap-strain
relationship. Another interesting prediction made by the Yang-Han
theory of CNTs electromechanical response is periodic oscillations of
the bandgap with the applied strain, as the shifted Fermi points cross
different allowed sub-bands. Evidence of such oscillations can be seen
in the upper- and lower-left panels of Fig.~\ref{Fig: GapBending}. Due
to the relatively small dimensions of the GNRs studied, only a partial
period is obtained within the deformation depth range studied. This
resembles recent measurements made on CNTs, where the frequency of the
bandgap oscillations was found to decrease with increasing
diameter.~\cite{Nagapriya2008}

\begin{figure}[h]
\begin{center}
\epsfxsize=14.0cm \epsffile{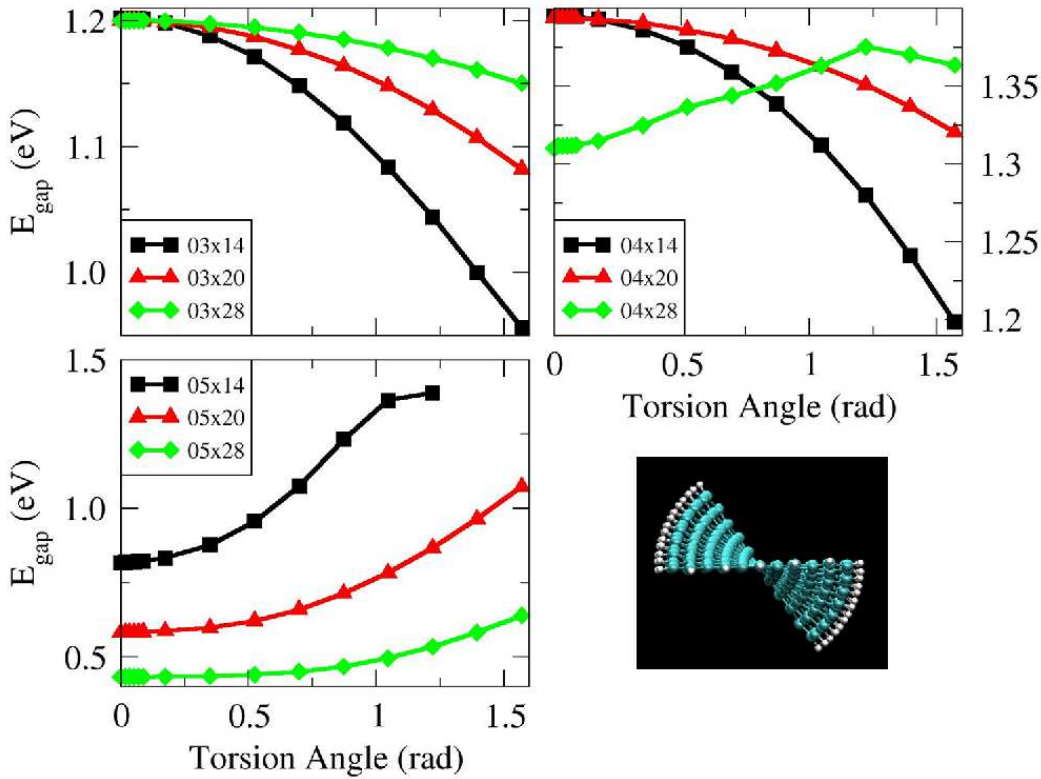}
\end{center}
\caption{HOMO-LUMO gap changes of the $03$x$N$ (upper left panel), $04$x$N$
(upper right panel), and $05$x$N$ (lower left panel) suspended
graphene nanoribbons due to an externally applied torsional stress.}
\label{Fig: GapTwisting}
\end{figure}

A very similar picture arises for the case of torsional
deformations. Fig.~\ref{Fig: GapTwisting} presents the dependence of
the energy gap on the torsional angle, up to a value of
$\phi=90^o$. As for the case of bending, energy gap changes of up to
$0.5$ eV are obtained upon torsion of the smaller systems studied.
The sensitivity reduces as the dimensions of the systems are
increased, and the general trend of increase (decrease) in the energy
gap of the $03$x$N$, $04$x$N$ ($05$x$N$), upon the appearance of
stresses in the system, remains. Remnants of the energy gap
oscillations can be seen for the $05$x$14$ nanoribbon in the lower
left panel of the figure.
\input{epsf}

%% file: Summary.tex
In summary, we have studied the electromechanical properties of
suspended graphene nanoribbons under bending and torsional
deformations. High sensitivity of the electronic properties to applied
stresses was found, suggesting their potential use as building blocks
in nanoelectromechanical devices.  Classical elasticity theory
adequately describes the mechanical behavior of the systems under
bending deformations.  The calculated Young's modulus of $7$ TPa marks
ultra-narrow GNRs as one of the strongest existing materials.  While
the systems we study are probably too small to be manipulated by an
external tip as illustrated in Fig.~\ref{Fig: Ribbon}, the results of
our calculations present important trends that are expected to hold
for larger systems as well. An alternative platform to induce bending
deformations on molecular graphene derivatives~\cite{Yang2008} such as
those studied herein, while measuring their transport properties would
be the use of mechanically controllable break
junctions.~\cite{Agrait2003, Selzer2006} In such a setup a graphene
ribbon bridging the gap of a pre-designed molecular scale junction may
be subject to delicate bending deformations via the careful
manipulation of the underlying surface. With this respect, an
interesting experimental challenge would be to induce, in a similar
manner, torsional deformations of graphene nanoribbons.



\section{Acknowledgments}
This work was supported by NSF Award Number CHE-0807194 and the Welch
Foundation.  Work in Israel was supported by the Israel Science
Foundation (Grant Number 1313/08). Calculations were performed in part
on the Rice Terascale Cluster funded by NSF under Grant EIA-0216467,
Intel, and HP, and on the Shared University Grid at Rice funded by NSF
under Grant EIA-0216467, and a partnership between Rice University,
Sun Microsystems, and Sigma Solutions, Inc. O.H. would like to thank
Prof. Boris I. Yakobson for insightful discussions on the subject.
